\documentclass[12pt,a4paper]{article}

\usepackage{fourier}
\usepackage{microtype}
\usepackage[hmargin=2.5cm, vmargin=3.0cm]{geometry}
\usepackage{graphicx}
\usepackage{amssymb,amsmath}
\usepackage{dsfont}
\usepackage[official]{eurosym}
\usepackage{natbib}
\usepackage[pagebackref=true,
            colorlinks=true,
            citecolor=blue  ]{hyperref} 

\numberwithin{equation}{section}

\newcommand{\Dt}{\Delta t}
\newcommand{\DD}{\mathds{D}}  
\newcommand{\e}{\text{\euro}} 
\newcommand{\EE}{\mathds{E}}  
\newcommand{\EED}{\mathds{E}^\DD}  
\newcommand{\EES}{\mathds{E}^\QS}  
\newcommand{\eps}{\varepsilon}
\newcommand{\half}{\mathchoice
 {{\textstyle{\frac{1}{2}}}}
 {{\textstyle{\frac{1}{2}}}}
 {{\scriptscriptstyle{\frac{1}{2}}}}
 {{\scriptscriptstyle{\frac{1}{2}}}}
}
\newcommand{\hp}{h^{\text{p}}}
\newcommand{\hv}{h^{\text{v}}}
\newcommand{\Oh}{\mathcal{O}}
\newcommand{\pic}{\pi^\text{c}} 
\newcommand{\piD}{\pi^\text{D}} 
\newcommand{\pim}{\pi^\text{m}} 
\newcommand{\pimf}{\pi^\text{mf}} 
\newcommand{\pip}{\pi^\text{p}} 
\newcommand{\pis}{\pi^\text{s}} 
\newcommand{\piv}{\pi^\text{v}} 
\newcommand{\PP}{\mathds{P}}  
\newcommand{\QS}{\mathds{S}}  
\newcommand{\Var}{\mathds{V}\kern-2pt\text{ar}} 
\newcommand{\VaR}{\text{VaR}} 

\begin{document}

\title{Time-Consistent Actuarial Valuations}

\author{Antoon Pelsser%
\thanks{The author would like to thank Monique Jeanblanc, Dilip Madan, Eckhard Platen, Michel Vellekoop, participants at the Actuarial \& Financial Mathematics 2010 conference in Brussels and the Bachelier Seminar in Paris for comments and helpful suggestions.}\\
\\
Maastricht University \& Netspar\\
Dept.~of Quantitative Economics and Dept.~of Finance\\
P.O.~Box 616\\
6200 MD Maastricht\\
The Netherlands\\
Email:~a.pelsser@maastrichtuniversity.nl}

\date{%
First version: November 28, 2009\\
This version: \today}

\maketitle

\begin{abstract}
\noindent Recent theoretical results establish that time-consistent valuations (i.e.~pricing operators) can be created by backward iteration of one-period valuations. In this paper we investigate the continuous-time limits of well-known actuarial premium principles when such backward iteration procedures are applied. We show that the one-period variance premium principle converges to the non-linear exponential indifference valuation. Furthermore, we study the convergence of the one-period standard-deviation principle and establish that the Cost-of-Capital principle, which is widely used by the insurance industry, converges to the same limit as the standard-deviation principle. Finally, we study the connections between our time-consistent pricing operators, Good Deal Bound pricing and pricing under model ambiguity.
\end{abstract}

\pagebreak
\section{Introduction}\label{sec:Intr}

Standard actuarial premium principles usually consider a static premium calculation problem: what is the price today of an insurance contract with payoff at time $T$. See, for example, the textbooks by \citet{buhlmann1970mathematical}, \citet{gerber1979introduction}, or \citet{Kaas:MART}. Also, the study of risk measures, and the closely related concept of monetary risk measures has been studied in such a static setting. See, for example, \citet{ADEH:1999coherent}, \citet{cheridito2005coherent}. Also the study of utility indifference valuations has mainly confined itself to this static setting. For different applications we mention a few papers: \citet{Henderson:02}, \citet{Young:02}, \citet{hobson2004q-optimal}, \citet{Musiela:04}, \citet{monoyios2006characterisation}, and the recent book by \citet{carmona2009indifference}.

Financial pricing usually considers a ``dynamic'' pricing problem: how does the price evolve over time until the final payoff date $T$. This dynamic perspective is driven by the focus on hedging and replication. This literature started by the seminal paper of \citet{bs:optprice} and has been immensely generalised to broad classes of securities and stochastic processes, see \citet{Delbaen:Schachermayer:94}.

In recent years, researchers have begun to investigate risk measures in a dynamic setting, where the question of constructing time-consistent (or ``dynamic'') risk-measures has been investigated. See, \citet{riedel2004dynamic:coherent}, \citet{cher:delb:kupp:2006:dynamic}, \citet{roorda2005coherent}, \citet{rosazza2006risk-g-expect}, \citet{ADEHK2007coherent}. In a recent paper \citet{jobert:rogers:2008:valuations} show how time-consistent valuations can be constructed via backward induction of static one-period risk-measures (or ``valuations'').

In this paper we want to investigate well-known actuarial premium principles such as the one-period variance principle and the standard-deviation principle, and study their time-consistent extension. The method we use to construct these extensions is to first consider one-period valuations, then extend this to a multi-period setting using the backward iteration method of \citet{jobert:rogers:2008:valuations} for a given discrete time-step $\Dt$, and finally consider the continuous-time limit for $\Dt\to0$. We show that the one-period variance premium principle converges to the non-linear exponential indifference valuation. Furthermore, we study the convergence of the one-period standard-deviation principle and establish that the Cost-of-Capital principle, which is widely used by the insurance industry, converges to the same limit as the standard-deviation principle.

The rest of this paper is organised as follows. We start in Section~\ref{sec:VarPric} where we focus initially on the case of pure insurance risk. This allows us to clearly demonstrate the construction we use to derive the continuous-time limit of time-consistent actuarial pricing principles. We derive the limit for the Variance principle, the Standard-Deviation principle and the Cost-of-Capital principle. \dots We summarise and conclude in Section~\ref{sec:Concl}.

\section{Variance Pricing}\label{sec:VarPric}

\subsection{Diffusion Setting}\label{sec:Diff}

We start by considering an unhedgeable insurance process $y_t$, which is given by a diffusion equation:
\begin{equation}\label{eq:defy}
  dy = a(t,y)\,dt + b(t,y)\,dW.
\end{equation}
We also consider a discretisation scheme for the insurance process in the form of a binomial tree:
\begin{equation}\label{eq:binoy}
  y(t+\Dt) = y(t) + a\Dt + \left\{
  \begin{array}{ll}
    + b\sqrt{\Dt} & \text{with prob.~}\half\\
    - b\sqrt{\Dt} & \text{with prob.~}\half\\
  \end{array}\right.
\end{equation}
where we have suppressed the dependence of $a$ and $b$ on $(t,y)$ to lighten the notation. Note that we have restricted ourselves to a Markovian diffusion setting, which allows us to give a very simple mathematical derivation of our results. In Section~\ref{sec:BSDE} we discuss alternatives for this restrictive assumption.

Given the discrete-time setting \eqref{eq:binoy}, we can now create time-consistent pricing operators (``valuations''), using the backward induction method of \citet{jobert:rogers:2008:valuations}. Let us denote a one-step valuation by $\Pi[]$, and the resulting price at $(t,y)$ by $\pi(t,y)$:
\begin{equation}\label{eq:pi-onestep}
  \pi(t,y) = \Pi_t\bigl[\pi(t+\Dt,y(t+\Dt))\bigr]
\end{equation}
In words: the price $\pi(t,y)$ is obtained by applying at time $t$ the one-step valuation $\Pi_t[]$ to the random variable $\pi(t+\Dt,y(t+\Dt))$, which is the price obtained in the previous time-step.

\subsection{Variance Principle}\label{sec:VarPrin}

If we consider an insurance contract with a payoff at time $T$ defined as a function $f\bigl(y(T)\bigr)$, then the actuarial Variance Principle $\Pi^{\text{v}}_t[]$ is defined as \citep[see, e.g.{}][]{Kaas:MART}
\begin{equation}\label{eq:defVarPri}
  \Pi^{\text{v}}_t[f(y(T))] = \EE_t[f(y(T))] + \half\alpha\Var_t[f(y(T))],
\end{equation}
where $\EE_t[]$ and $\Var_t[]$ denote the expectation and variance operators conditional on the information available at time $t$ under the ``real-world'' probability measure $\PP$.

Note that in the standard actuarial literature \citep[see, e.g.{}][]{Kaas:MART}, discounting is usually ignored. To facilitate the discussion, we will first derive the continuous-time limit of the variance principle without using discounting in Section~\ref{sec:VarPrin:NoDis}. We will then consider case with discounting in Section~\ref{sec:VarPrin:WithDis}, and discuss further generalisations in Section~\ref{sec:BSDE}.

\subsubsection{No Discounting}\label{sec:VarPrin:NoDis}

In the binomial tree discretisation we can obtain an explicit expression for a one-step variance price $\piv(t,y)$ by substituting \eqref{eq:defVarPri} into \eqref{eq:pi-onestep}:
\begin{equation}\label{eq:bino-piv}
  \piv\bigl(t,y(t)\bigr) =\EE_t[\piv\bigl(t+\Dt,y(t+\Dt)\bigr)] + \half\alpha\Var_t[\piv\bigl(t+\Dt,y(t+\Dt)\bigl)].
\end{equation}
We are now interested in considering the limit for $\Dt\to0$. We assume that $\piv(t+\Dt, y)$ is sufficiently smooth to be twice continuously differentiable in~$y$, such that we can apply for all values of $y$ the Taylor approximation
\begin{multline}\label{eq:piv-taylor}
  \piv\bigl(t+\Dt, y(t)+h\bigr) = \piv\bigl(t+\Dt, y(t)\bigr) + {} \\ \piv_y\bigl(t+\Dt, y(t)\bigr)h + \half\piv_{yy}\bigl(t+\Dt, y(t)\bigr)h^2 + \Oh(h^3),
\end{multline}
where subscripts on $\piv$ denote partial derivatives. If we substitute this Taylor approximation for the binomial approximation \eqref{eq:binoy} into \eqref{eq:bino-piv} and gather all terms in ascending orders of $\Dt$, we obtain
\begin{multline}\label{eq:bino-piv-tay1}
  \piv\bigl(t,y(t)\bigr) - \piv\bigl(t+\Dt,y(t)\bigr) =
  a\piv_y\bigl(t+\Dt,y(t)\bigr)\Dt +{} \\
  \half b^2\piv_{yy}\bigl(t+\Dt,y(t)\bigr)\Dt +
  \half\alpha \Bigl(b\piv_y\bigl(t+\Dt,y(t)\bigr)\Bigr)^2\Dt +
  \Oh(\Dt^2).
\end{multline}
If we divide by $\Dt$ we obtain
\begin{multline}\label{eq:bino-piv-tay2}
  \frac{\piv\bigl(t,y(t)\bigr) - \piv\bigl(t+\Dt,y(t)\bigr)}{\Dt} =
  a\piv_y\bigl(t+\Dt,y(t)\bigr) +{} \\
  \half b^2\piv_{yy}\bigl(t+\Dt,y(t)\bigr) +
  \half\alpha \Bigl(b\piv_y\bigl(t+\Dt,y(t)\bigr)\Bigr)^2 +
  \Oh(\Dt).
\end{multline}
We can now take the limit for $\Dt\to0$. The left-hand side of \eqref{eq:bino-piv-tay2} converges to (minus) the partial derivative of $\piv()$ with respect to $t$ (i.e.~$-\piv_t\bigl(t,y(t)\bigr)$), and we obtain
\begin{equation}\label{eq:pde-piv}
  \piv_t +
  a\piv_y +
  \half b^2\piv_{yy} +
  \half\alpha (b\piv_y)^2 = 0,
\end{equation}
where we have suppressed (again) the dependence on $t$ and $y$ to lighten the notation.

Note, that equation \eqref{eq:pde-piv} is a semi-linear partial differential equation (pde) that describes the behaviour of the variance price $\piv(t,y)$ as a function of $t$ and $y$. The pde is subject to the boundary condition $\piv\bigl(T,y(T)\bigr) = f\bigl(y(T)\bigr)$ which is the payoff of the insurance contract at time~$T$. The existence of solutions of semi-linear pde's has been studied in the context of Backward Stochastic Differential Equations (BSDE's), and we discuss this subject further in Section~\ref{sec:BSDE}.

In this particular case, we can construct the solution of \eqref{eq:pde-piv} explicitly by employing a Hopf-Cole transformation of the solution that removes the non-linearity from the pde. Consider the auxiliary function $\hv(t,y) := \exp\{\alpha \piv(t,y)\}$. The original function $\piv(t,y)$ can be obtained from the inverse relation $\piv(t,y) = \frac{1}\alpha\ln{\hv(t,y)}$. If we now apply the chain-rule of differentiation, we can express the partial derivatives of $\piv()$ in terms of $\hv()$ as
\begin{equation}\label{eq:piv-hv-ders}
    \piv_t = \frac{1}{\alpha}\frac{\hv_t}{\hv},\quad
    \piv_y = \frac{1}{\alpha}\frac{\hv_y}{\hv},\quad
    \piv_{yy} = \frac{1}{\alpha}\frac{\hv_{yy}\hv - (\hv_y)^2}{(\hv)^2}.
\end{equation}
If we substitute these expressions into \eqref{eq:pde-piv}, the non-linear terms cancel and we obtain a linear pde for $\hv(t,y)$:
\begin{equation}\label{eq:pde-hv}
  \hv_t + a\hv_y + \half b^2 \hv_{yy}=0.
\end{equation}
Hence, by considering the transformed function $\hv(t,y)$ we have managed to obtain a linear pde for $\hv()$. The boundary condition at $T$ is given by $\hv(T,y(T)) = \exp\{\alpha \piv(T,y(T))\} = \exp\{\alpha f(y(T))\}$. Using the Feynman-Ka\c c formula, we can express the solution of \eqref{eq:pde-hv} as
\begin{equation}\label{eq:hv-FK}
  \hv(t,y) = \EE_t\left[ e^{\alpha f(y(T))} \middle\vert y(t)=y \right],
\end{equation}
where the expectation is taken with respect to the stochastic process $y(t)$ defined in equation~\eqref{eq:defy} conditional on the information that at time $t$ the process $y(t)$ is equal to $y$. From the representation \eqref{eq:hv-FK} follows immediately that we can express $\piv(t,y)$ as
\begin{equation}\label{eq:piv-FK}
  \piv(t,y) = \frac{1}{\alpha} \ln\EE_t\left[ e^{\alpha f(y(T))} \middle\vert y(t)=y \right].
\end{equation}
Note that this representation of the variance-price $\piv()$ is identical to the exponential indifference price which has been extensively studied in recent years. See, for example \citet{Henderson:02}, \citet{Young:02} or \citet{Musiela:04}. For an overview of recent advances in indifference pricing, we refer to the book by \citet{carmona2009indifference}.

To summarise this section, we have established that the continuous-time limit of the iterated actuarial variance principle is the exponential indifference price.

\subsubsection{With Discounting}\label{sec:VarPrin:WithDis}

Up to now we have ignored discounting in our derivation. (Or equivalently, we assumed that the interest rate is equal to zero.) In a time-consistent setting, it is important to take discounting into consideration, as money today cannot be compared to money tomorrow.

If we consider the definition of the variance principle given in~\eqref{eq:defVarPri}, it seems that we are adding apples and oranges together. The first term $\EE_t[f(y(T))]$ is a quantity in monetary units (say \e) at time $T$. However, the second term $\Var_t[f(y(T))]$ is basically the expectation of $f(y(T))^2$, and is therefore a quantity in units of $(\!\e)^2$. The way to rectify this situation is by understanding that the parameter $\alpha$ is not a dimensionless quantity, but is a quantity expressed in units of $1/\!\e$. This should not come as a surprise. The parameter $\alpha$ is in fact the absolute risk aversion parameter introduced by seminal paper by \citet{pratt:1964:risk} where he derives the variance principle as an approximation ``in the small'' of the price that an economic agent facing a decision under uncertainty should ask.

To stress in our notation the units in which the absolute risk aversion $\alpha$ is expressed, we will rewrite the absolute risk aversion as the relative risk aversion $\gamma$ \citep[also introduced by][]{pratt:1964:risk}, which is a dimensionless quantity, divided by a benchmark wealth-level $X(T)$, which is expressed in \e~at time~$T$. If we now assume a constant rate of interest $r$, we can then set our benchmark wealth as $X(T) = X_0 e^{rT}$. Hence, we rewrite our variance principle as
\begin{equation}\label{eq:defVarPri2}
  \Pi^{\text{v}}_t[f(y(T))] =
  \EE_t[f(y(T))] + \half\frac{\gamma}{X_0 e^{rT}}\Var_t[f(y(T))].
\end{equation}
Note, that $\Pi^{\text{v}}_t[]$ leads to a ``forward'' price expressed in units of $\e$ at time~$T$.

Given the enhanced definition \eqref{eq:defVarPri2} of the variance principle including discounting, we can now proceed as in Section~\ref{sec:VarPrin:NoDis}. For a single binomial step, we obtain the following expression for the price:
\begin{multline}\label{eq:bino-pivd}
  \piv\bigl(t,y(t)\bigr) =
  e^{-r\Dt}\bigg(
  \EE_t[\piv\bigl(t+\Dt,y(t+\Dt)\bigr)] + \\
  \half\frac{\gamma}{X_0 e^{r(t+\Dt)}}\Var_t[\piv\bigl(t+\Dt,y(t+\Dt)\bigl)]\bigg).
\end{multline}
Note that we have included an additional discounting term $e^{-r\Dt}$ to discount the values from time $t+\Dt$ back to time $t$. Using a similar derivation as before, we arrive at the following partial differential equation for $\piv(t,y)$:
\begin{equation}\label{eq:pde-pivd}
  \piv_t +
  a\piv_y +
  \half b^2\piv_{yy} +
  \half\frac{\gamma}{X_0 e^{rt}} (b\piv_y)^2 - r\piv = 0.
\end{equation}
This non-linear pde can again be linearised by considering the transformation $\hv(t,y) = \exp\{\frac{\gamma}{X_0 e^{rt}}\piv(t,y)\}$, which leads to the following expression for the solution of~\eqref{eq:pde-pivd}:
\begin{equation}\label{eq:pivd-FK}
  \piv(t,y) = \frac{X_0 e^{rt}}{\gamma}
  \ln\EE\left[ e^{\frac{\gamma}{X_0 e^{rT}} f(y(T))} \middle\vert y(t)=y \right].
\end{equation}
From this result we see that the discounting is incorporated into the non-linear pricing formula, by expressing all units relative to the ``benchmark wealth'' $X(t)=X_0 e^{rt}$.\footnote{For general results concerning ``benchmark pricing'' in a linear setting we refer to \citet{platen2006benchmark} and the book by \citet{PlatenHeath2006benchmark}.} See chapter by MuZa in Carmona book??

\subsubsection{Current price as benchmark}\label{sec:pipPrin}

In the previous subsection we have taken the benchmark wealth to be a risk-free investment $X_0 e^{rt}$. Another interesting example is when we consider the current price $\pi(t,y)$ as the benchmark wealth.

This then leads to a new pricing operator, which we will denote by $\pip()$. The one-step valuation is then given by
\begin{equation}\label{eq:bino-pip}
  \pip\bigl(t,y(t)\bigr) =
  e^{-r\Dt}\bigg(
  \EE_t[\pip\bigl(t+\Dt,y(t+\Dt)\bigr)] +
  \half\gamma\frac{\Var_t[\pip\bigl(t+\Dt,y(t+\Dt)\bigl)]}
  {\EE_t[\pip\bigl(t+\Dt,y(t+\Dt)\bigr)]}
  \bigg).
\end{equation}
Hence, we assume that we want to measure the variance of $\pip()$ relative to the expected value of $\pip()$. Obviously, this is only well-defined if $\pip(t,y)$ is strictly positive for all $(t,y)$.

If we employ our Taylor-expansion and take the limit for $\Dt\to0$ we obtain the following pde
\begin{equation}\label{eq:pde-pip}
  \pip_t +
  a\pip_y +
  \half b^2\pip_{yy} +
  \half\frac{\gamma}{\pip} (b\pip_y)^2 - r\pip = 0.
\end{equation}
Again, we can study the solution of \eqref{eq:pde-pip} by employing a transformation of the solution that removes the non-linearity from the pde. Consider the auxiliary function $\hp(t,y) := (\pip(t,y))^{1/q}$. The original function can be obtained from the inverse relation $\pip(t,y) = (\hp(t,y))^q$. If we now apply the chain-rule of differentiation, we can express the partial derivatives of $\pip()$ in terms of $\hp()$ as
\begin{equation}\label{eq:pip-hp-ders}
    \pip_t = q(\hp)^{q-1}\hp_t,\quad
    \pip_y = q(\hp)^{q-1}\hp_y,\quad
    \pip_{yy} = q(\hp)^{q-1}
      \left(\frac{q-1}{\hp}(\hp_y)^2+\hp_{yy}\right).
\end{equation}
If we substitute these expressions into \eqref{eq:pde-pip} and simplify, we obtain
\begin{equation}
  \hp_t + a \hp_y +
  \half b^2
  \left(\frac{(1+\gamma)q-1}{\hp}(\hp_y)^2+\hp_{yy}\right) - \frac{r}{q}\hp = 0.
\end{equation}
If we choose $q=1/(1+\gamma)$, then
the non-linear terms cancel and we obtain a linear pde for $\hp(t,y)$:
\begin{equation}\label{eq:pde-hp}
  \hp_t + a\hp_y + \half b^2 \hp_{yy} - r(1+\gamma)\hp =0.
\end{equation}
The boundary condition at $T$ is given by $\hp(T,y(T)) = \piv(T,y(T))^{1+\gamma} = f(y(T))^{1+\gamma}$. Using the Feynman-Ka\c c formula, we can express the solution of \eqref{eq:pde-hp} as
\begin{equation}\label{eq:hp-FK}
  \hp(t,y) = \EE_t\left[ e^{-r(1+\gamma)(T-t)}f(y(T))^{1+\gamma} \middle\vert y(t)=y \right],
\end{equation}
where the expectation is taken with respect to the stochastic process $y(t)$ defined in equation~\eqref{eq:defy} conditional on the information that at time $t$ the process $y(t)$ is equal to $y$. From the representation \eqref{eq:hp-FK} follows immediately that we can express $\pip(t,y)$ as
\begin{equation}\label{eq:pip-FK}
  \pip(t,y) = e^{-r(T-t)}\left(\EE_t\left[ f(y(T))^{1+\gamma} \middle\vert y(t)=y \right]\right)^\frac{1}{1+\gamma},
\end{equation}
Note that this representation of the price $\pip()$ arises also in the study of indifference pricing under power-utility functions, and the related notion of pricing under so-called ``$q$-optimal'' measures. See, for example \citet{hobson2004q-optimal} and \citet{henderson:hobson:2009:carmona}.

\subsection{Mean Value Principle}\label{sec:MVprin}

The examples we gave in the previous subsections, are all special cases of the \emph{Mean Value Principle}, which is defined as
\begin{equation}\label{eq:defMVPri}
  \Pi^{\text{m}}_t[f(y(T))] =
  v^{-1}\left(\EE_t[v(f(y(T)))]\right)
\end{equation}
for any function $v()$ which is a convex and increasing \citep[see][Chap.~5]{Kaas:MART}.

Once more, we have to pay attention to units. If we want to apply a general function $v()$ to a value (expressed in units of $\e$), we have to make sure that the argument of $v()$ is dimensionless. The easiest way to achieve this, is to express the argument for $v()$ in ``forward terms''. For a single binomial step, we therefore obtain the following expression for the price:
\begin{equation}\label{eq:bino-pim}
  \frac{\pim\bigl(t,y(t)\bigr)}{e^{rt}} =
  v^{-1}\!\left(
  \EE_t\left[
  v\!\left(
  \frac{\pim\bigl(t+\Dt,y(t+\Dt)\bigr)}
  {e^{r(t+\Dt)}}
  \right)
  \right]
  \right).
\end{equation}
We can rewrite this definition as
\begin{equation}\label{eq:bino-pim-2}
  v\!\left(
  \frac{\pim\bigl(t,y(t)\bigr)}{e^{rt}}
  \right) =
  \EE_t\left[
  v\!\left(
  \frac{\pim\bigl(t+\Dt,y(t+\Dt)\bigr)}
  {e^{r(t+\Dt)}}
  \right)
  \right],
\end{equation}
and from this expression it is immediately obvious that the ``distorted'' value $v(\pim(t,y)/e^{rt})$ is linear and therefore satisfies the Feynman-Ka\c c formula. Which corresponds exactly to the solutions we found in the previous subsections.

In this case, we want to go in the opposite direction and we seek the corresponding pde for the price $\pim(t,y)$. To do this, we will use our Taylor-expansion derivation, but with an additional twist: we must expand the functions $v()$ and $v^{-1}()$ as well. In particular, we seek to expand the function $v^{-1}(v(x)+h)$ for small $h$. Using the identity $v^{-1}(v(x))\equiv x$ we obtain
\begin{equation}
v^{-1}(v(x)+h) = x + \frac{h}{v'(x)} - \frac{v''(x)}{2(v'(x))^3}h^2 + \Oh(h^3).
\end{equation}
Combining this result with the Taylor expansions for $v()$ and $\pim()$ on a single binomial time-step, and taking the limit for $\Dt\to0$ leads to the pde:
\begin{equation}\label{eq:pde-pimf}
  \pimf_t +
  a\pimf_y +
  \half b^2\pimf_{yy} +
  \half\frac{v''(\pimf)}{v'(\pimf)} (b\pimf_y)^2 = 0,
\end{equation}
where $\pimf(t,y) := \pim(t,y)/e^{rt}$ is the price expressed in forward terms.

We see in equation \eqref{eq:pde-pimf}, that the coefficient in front of the non-linear term can be identified as the ``local risk aversion'' induced by the function $v()$ at the current value $\pimf()$. Note, that since the function $v()$ is increasing and convex by assumption, $v''()/v'()$ is positive.

\section{Standard-Deviation Pricing}\label{sec:StDev}

\subsection{Standard-Deviation Principle}\label{sec:StDevPrin}

Another well-known actuarial pricing principle is the Standard Deviation Principle, \citep[see][]{Kaas:MART} defined as
\begin{equation}\label{eq:defStDPri}
  \Pi^{\text{s}}_t[f(y(T))] = \EE_t[f(y(T))] + \beta\sqrt{\Var_t[f(y(T))]}.
\end{equation}
Please note that also in this case we have to be careful about the dimensionality of the parameter $\beta$. Even though both the expectation and the standard deviation are expressed in units of $\e$, the standard deviation and the expectation have different ``time-scales''. If we go down to small time-scales (as we will be doing when considering the limit for $\Dt\to0$) then due to the diffusion term $dW$ of the process $y$, we have the property that the expectation of any function $f(y)$ scales linearly with $\Dt$, but the standard deviation scales with $\sqrt{\Dt}$. This means that for small $\Dt$ the standard deviation term will completely overpower (literally!) the expectation term. Therefore, the only way to obtain a well-defined limit for $\Dt\to0$ is if we take $\beta\sqrt{\Dt}$ as the parameter for the standard deviation principle for a binomial step.

Another way of understanding this result, is to consider the following example. If we want to compare a standard deviation measured over an annual time-step with a standard deviation measured over a monthly time-step, we have to scale the annual outcome with $\sqrt{1/12}$ to get a fair comparison.

Given the above discussion on the time scales, we get for a single binomial step, the following expression for the price:
\begin{equation}\label{eq:bino-pis}
  \pis\bigl(t,y(t)\bigr) =
  e^{-r\Dt}\bigg(
  \EE_t[\pis\bigl(t+\Dt,y(t+\Dt)\bigr)] +
  \beta\sqrt{\Dt}\sqrt{\Var_t[\pis\bigl(t+\Dt,y(t+\Dt)\bigl)]}
  \bigg).
\end{equation}
Using a similar derivation as in Section~\ref{sec:VarPrin}, we arrive at the following partial differential equation for $\pis(t,y)$:
\begin{equation}
  \pis_t +
  a\pis_y +
  \half b^2\pis_{yy} +
  \beta \sqrt{(b\pis_y)^2} - r\pis = 0,
\end{equation}
which can be rewritten as
\begin{equation}\label{eq:pde-pis}
  \pis_t +
  a\pis_y +
  \half b^2\pis_{yy} +
  \beta b|\pis_y| - r\pis = 0.
\end{equation}
This is again once more a non-linear pde. However, the non-linearity is much more benign in this case. Whenever the partial derivative $\pis_y$ does not change sign on the whole domain of $y$ (i.e.~the function $\pis$ is either monotonically increasing or monotonically decreasing in $y$), then \eqref{eq:pde-pis} reduces to the linear pde:
\begin{equation}\label{eq:pde-pis-lin}
  \pis_t +
  (a\pm\beta b)\pis_y +
  \half b^2\pis_{yy} - r\pis = 0,
\end{equation}
where the sign of $\pm\beta b$ depends on the (uniquely defined) sign of $\pis_y$.

Using the Feynman-Ka\c c formula, we can represent the solution to~\eqref{eq:pde-pis-lin} as:
\begin{equation}\label{eq:pis-FK}
  \pis(t,y) = \EES_t\left[ f(y(T)) \middle\vert y(t)=y \right],
\end{equation}
where $\EES_t[]$ denotes the expectation at time $t$ with respect to the ``risk-adjusted'' process $y$ defined as
\begin{equation}\label{eq:defy-ra}
  dy = \bigl(a(t,y)\pm\beta b(t,y)\bigr)\,dt + b(t,y)\,dW^\QS.
\end{equation}
The drift-rate is adjusted upwards ($a+\beta b$) if the payoff  $f(y)$ is monotonically increasing in $y$, and adjusted downwards ($a-\beta b$) if $f(y)$ is monotonically decreasing in $y$. So, the risk-adjustment is always in the ``upwind'' direction of the risk, thus making the price $\pis$ more expensive than the real-world expectation $\EE[f(y)]$.


\subsection{Cost-of-Capital Principle}\label{sec:CoCPrin}

Another actuarial pricing principle is the Cost-of-Capital Principle. This was introduced by the Swiss insurance supervisor as a part of the method to calculate solvency capitals for insurance companies \citep{SST:2004:whitepaper}.\footnote{For a critical discussion on the risk-measure implied by the Swiss Solvency Test we refer to \citet{filipovic:2008:noteSST}.} The Cost-of-Capital method has been widely adopted by the insurance industry in Europe, and has also been prescribed as the standard method by the European Insurance and Pensions Supervisor for the Quantitative Impact Studies \citep[see][]{ceiops-QIS5}.

The Cost-of-Capital is based on the following economic reasoning. We first consider the ``expected loss'' $\EE[f(y(T)]$ of the insurance claim $f(y(T))$ as a basis for pricing. But this is not enough, the insurance company also has to hold a capital buffer against the ``unexpected loss''. This buffer is calculated as a Value-at-Risk over a time-horizon (typically 1 year) and a probability threshold $q$ (usually 0.995, or even higher). The unexpected loss is then calculated as $\VaR_q\bigl[f(y(T)) - \EE[f(y(T))]\bigr]$. The capital buffer is borrowed from the shareholders of the insurance company (i.e.~the buffer is subtracted from the surplus in the balance sheet). Given the very high confidence level, in most cases the buffer can be returned to the shareholders, however there is a small probability $(1-q)$ that the capital buffer is needed to cover an unexpected loss. Hence, the shareholders require a compensation for this risk in the form of a ``cost-of-capital''. This cost-of-capital needs to be included in the pricing of the insurance contract. If we denote the cost-of-capital by $\delta$, then the Cost-of-Capital Principle is given by
\begin{equation}  \label{eq:defCoCPri}
  \Pi^{\text{c}}_t[f(y(T))] = \EE_t[f(y(T))] +
  \delta\VaR_{q,t}\Bigl[f(y(T))-\EE_t[f(y(T))]\Bigr].
\end{equation}
Please note that also in this case we have to be careful about the dimensionality of the different terms. First, we are comparing Value-at-Risk quantities at different time-scales, and these have to be scaled back to a per annum basis, to do this we divide the $\VaR$-term by $\sqrt{\Dt}$. Then, we must realise that the cost-of-capital $\delta$ behaves like an interest rate: it is the compensation the insurance company needs to pay to its shareholders for borrowing the buffer capital over a certain period. The cost-of-capital is expressed as a percentage per annum, hence over a time-step $\Dt$ the insurance company has to pay a compensation of $\delta\Dt$ per $\e$ of buffer capital. As a result, we obtain a ``net scaling'' of $\delta\Dt / \sqrt{\Dt} = \delta\sqrt{\Dt}$. Note, that this is the same scaling as for the standard deviation principle.

For a single time-step, we therefore get the following expression for the cost-of-capital price:
\begin{multline}\label{eq:step-pic}
  \pic\bigl(t,y(t)\bigr) =
  e^{-r\Dt}\bigg(
  \EE_t[\pic\bigl(t+\Dt,y(t+\Dt)\bigr)] + \\
  \delta\sqrt{\Dt}\VaR_{q,t}\Bigl[\pic\bigl(t+\Dt,y(t+\Dt)\bigr)-
  \EE_t[\pic\bigl(t+\Dt,y(t+\Dt)\bigr)\bigr]\Bigr]
  \bigg).
\end{multline}

In the previous sections we used a binomial discretisation of the process~$y$. However, in this case we have to be a bit more careful. Since we are considering a $(1-q)$-quantile with very small probability $(1-q)$, a simple binomial tree approximation is too crude to obtain an accurate representation of the $(1-q)$-quantile of the process $y$. We therefore consider a ``quadrinomial'' tree, where we make sure that we match the mean, the variance and the $(1-q)$-quantile of the process $y$ over a $\Dt$ time-step. We can do this, for example, by considering the following discretisation:
\begin{align}
  y(t+\Dt) = y(t) + a\Dt + &\left\{
  \begin{array}{ll}
    + kb\sqrt{\Dt} & \text{with prob.~}(1-q)\\
    + lb\sqrt{\Dt} & \text{with prob.~}\half-(1-q)\\
    - lb\sqrt{\Dt} & \text{with prob.~}\half-(1-q)\\
    - kb\sqrt{\Dt} & \text{with prob.~}(1-q)\\
  \end{array}\right.\label{eq:quady}\\
  &\phantom{x}\quad\text{where~} l =\sqrt{\frac{\half-(1-q)k^2}{\half-(1-q)}},\notag
\end{align}
and where $k$ is chosen such that $kb\sqrt{\Dt}$ matches the $(1-q)$-quantile of the random variable $y(t+\Dt)-\EE_t[y(t+\Dt)]$ for the time-step $\Dt$. In particular, the distribution for $y(t+\Dt)$ will be very close to a Gaussian for small $\Dt$, and we can compute $k$ from the Gaussian distribution function $\Phi()$ as $k = \Phi^{-1}(q)$. For example, if $q=0.995$, then $k= 2.58$ and $l = 0.971$ (rounded to three significant digits).

Given the quadrinomial discretisation~\eqref{eq:quady}, we can now proceed as in the previous sections, and derive a pde for the price operator $\pic(t,y)$:
\begin{equation}\label{eq:pde-pic}
  \pic_t +
  a\pic_y +
  \half b^2\pic_{yy} +
  \delta k b|\pic_y| - r\pic = 0.
\end{equation}
This pde is exactly the same as \eqref{eq:pde-pis}, except for the factor $\delta k$ instead of $\beta$ in front of $b|\pic_y|$. Of course, this should not come as a surprise, since for a small time-step $\Dt$ the $(1-q)$-quantile of $y(t+\Dt)$ converges to $k$ times the standard deviation $b\sqrt{\Dt}$, and hence the cost-of-capital pricing operator $\pic()$ should converge to the standard deviation pricing operator $\pis()$ with $\beta=\delta k$.

If the payoff $f(y(T))$ is monotonous in $y(T)$, we can represent the cost-of-capital price $\pic(t,y)$ in the same way as the standard-deviation price~\eqref{eq:pis-FK} with respect to the ``risk-adjusted'' process $y$
\begin{equation}\label{eq:defy-ra-coc}
  dy = \bigl(a(t,y)\pm\delta k b(t,y)\bigr)\,dt + b(t,y)\,dW.
\end{equation}

\subsection{Davis Price}

So far, we have established that the continuous-time limit of the iterated variance price $\piv()$ solves the non-linear pricing pde \eqref{eq:pde-pivd}. The continuous-time limit of the standard deviation principle $\pis()$ (and also the Cost-of-Capital principle) solves the linear pricing pde \eqref{eq:pde-pis-lin}, which is considerably easier to solve. In this section we provide a connection between these two pricing principles. We show that the linear standard deviation price can be interpreted as the ``small perturbation'' expansion of non-linear variance price. The core of this idea can be traced back to \citet{davis1997optionincomplete}.

Let us consider the small perturbation expansion in more detail. Suppose we already have an existing portfolio of insurance liabilities, where the variance price $\piv(t,y)$ has already been determined for all relevant $t$ and $y$. Consider now a small position in an additional insurance claim with payoff $\eps g\bigl(y(T)\bigr)$ at time $T$.  Let us assume that for small $\eps$ the total price of the insurance portfolio can be decomposed into $\piv(t,y) + \eps\piD(t,y)$, where $\piD()$ denotes the ``Davis-price'' of the additional claim $\eps g()$. The total price $\piv() + \eps\piD()$ should solve the pricing pde \eqref{eq:pde-pivd}, and we find
\begin{multline}
  (\piv_t+\eps\piD_t) +
  a(\piv_y+\eps\piD_y) +
  \half b^2(\piv_{yy}+\eps\piD_{yy}) + {}\\
  \half\frac{\gamma}{X_0 e^{rt}} b^2
  \left((\piv_y)^2+2\eps\piv_y\piD_y + \eps^2(\piD_y)^2\right) - r(\piv+\eps\piD) = 0.
\end{multline}
By definition, the price $\piv()$ solves the pde, and we simplify the expression to
\begin{equation}
  \piD_t +
  a\piD_y +
  \half b^2\piD_{yy} +
  \half\frac{\gamma}{X_0 e^{rt}} b^2
  \left(2\piv_y\piD_y + \eps(\piD_y)^2\right) -
  r\piD = 0.
\end{equation}
For small $\eps$, we can ignore the non-linear $\eps$-term, and we obtain a linear pde for the small perturbation Davis-price
\begin{equation}\label{eq:pde-davis}
  \piD_t +
  \left(a + \frac{\gamma}{X_0 e^{rt}} b^2\piv_y\right)\piD_y +
  \half b^2\piD_{yy} +
  r\piD = 0.
\end{equation}
Using the Feynman-Ka\c c formula, we can represent the solution to~\eqref{eq:pde-davis} as:
\begin{equation}\label{eq:davis-FK}
  \piD(t,y) = \EED_t\left[ f(y(T)) \middle\vert y(t)=y \right],
\end{equation}
where $\EED_t[]$ denotes the expectation at time $t$ with respect to the ``risk-adjusted'' process $y$ defined as
\begin{equation}\label{eq:defy-davis}
  dy = \left(a(t,y)+\frac{\gamma}{X_0 e^{rt}}b^2(t,y)\piv_y(t,y) \right)dt + b(t,y)\,dW^\DD.
\end{equation}
Note, that the Davis-price defined only ``relative'' to existing portfolio price $\piv()$. In particular, we can interpret the adjustment to the drift term as the risk-aversion $\gamma/X_0 e^{rt}$ times the local ``standard deviation'' of the existing price process $b(t,y)\piv_y(t,y)$ times the ``standard deviation'' $b(t,y)$ of the insurance process $y$ the drives the additional claim $g\bigl(y(T)\bigr)$.

\section{BSDE's and $g$-expectations}\label{sec:BSDE}

The non-linear pde's we have derived in \eqref{eq:pde-piv}, \eqref{eq:pde-pivd}, \eqref{eq:pde-pip} and \eqref{eq:pde-pimf} have been studied in recent years in the context of backward stochastic differential equations (or BSDE's). Necessary conditions for the existence and uniqueness of solutions of BSDE's have been established. For an overview of applications of BSDE's in finance, we refer to \citet{KarouiPengQuenez1997BSDE-Finance} and \citet{barrieu:elkaroui:2009:carmona}.

Using ``BSDE notation'' one can show that the solution to the pde  \eqref{eq:pde-pivd} can be represented by the triplet of processes $(y_t,Y_t,Z_t)$ satisfying
\begin{equation}\label{eq:BSDE}
\left\{
\begin{aligned}
  dy_t &= a(t,y_t)\,dt + b(t,y_t)\,dW_t\\
  dY_t &= -g(t,y_t,Y_t,Z_t)\,dt + Z_t\,dW_t\\
  Y_T &= f\bigl(y(T)\bigr),
\end{aligned}
\right.
\end{equation}
with ``generator'' $g(t,y,Y,Z) = \half\frac{\gamma}{X_0 e^{rt}} Z^2 -rY$. The realisation of the stochastic process $Y_t$ (depending on $y_t$) is then considered to be the solution to \eqref{eq:pde-pivd}. Another example: to represent the solution to \eqref{eq:pde-pimf} of the price under the mean value principle for a given function $v()$, we would use a BSDE with generator $g(t,y,Y,Z) = \half \bigl(v''(Y)/v'(Y)\bigr) Z^2$.

The mathematical setup of BSDE's allows significant generalisations over the restrictive Markovian diffusion setting we are using in this paper. For example, one can extend the setup to L\'evy processes, see \citet{Nualart:Schoutens:2001BSDE-Levy}. However, we will not explore that avenue in this paper.

Furthermore, solutions to BSDE's are always time-consistent, and can therefore be used to define non-linear ``$g$-expectations'' (where the prefix ``$g$'' refers to the generator $g(t,y,Y,Z)$ in \eqref{eq:BSDE}) and the related notion of ``$g$-martingales'', see, for example, \citet{peng2004filtration}. The connection between \hbox{$g$-expectations} and time-consistent risk measures has been explored in \citet{rosazza2006risk-g-expect}, \citet{barrieu:elkaroui:2009:carmona}, \citet{DelbPenRos:2010:Representationofpenalty}.

Link with coherent and convex risk measures, see RosGia.

%

\section{Good Deal Bound Pricing and Model Ambiguity}\label{sec:GDB}

In this section.

\section{Summary and Conclusions}\label{sec:Concl}
In this paper we have investigated well-known actuarial premium principles such as the variance principle and the standard-deviation principle, and studied their extension into both time-consistent and market-consistent directions. The method we used to construct these extensions was consider one-period valuations, then extend this to a multi-period setting using the backward iteration method of \citet{jobert:rogers:2008:valuations} for a given discrete time-step $\Dt$, and finally consider the continuous-time limit for $\Dt\to0$. We showed that the extended variance premium principle converges to the non-linear exponential indifference valuation. Furthermore, we showed that the extended standard-deviation principle converges to an expectation under an equivalent martingale measure. Finally, we showed that the Cost-of-Capital principle, which is widely used by the insurance industry, converges to the same limit as the standard-deviation principle.


\newpage
\bibliographystyle{apalike}
\bibliography{APbib}

\end{document}